\documentclass[useAMS,usenatbib]{mn2e}

\usepackage{graphicx}
\usepackage{amssymb}

\newcommand{\apj}{ApJ}

\newcommand{\aap}{A\&A}

\newcommand{\mnras}{MNRAS}

\newcommand{\kms}{\ensuremath{{\rm km}~{\rm s}^{-1}}}

\newcommand{\HI}{\ion{H}{i}}

\newcommand{\sunn}{$_{\odot}$}
\newcommand{\mr}{\ensuremath{{M_{\rm r}}}}
\newcommand{\mb}{\ensuremath{{M_{\rm B}}}}

\DeclareRobustCommand{\ion}[2]{%
\relax\ifmmode
\ifx\testbx\f
{\mathrm{#1\,\textsc{#2}}}\else
{\mathrm{#1\,\mathsc{#2}}}\fi
\else\textup{#1\,{\mdseries\textsc{#2}}}%
\fi}

\title[Extremely gas rich dwarf triplet in the Lynx-Cancer void] {Discovery of an extremely gas-rich dwarf triplet near the center of the Lynx-Cancer void}

\author[J.N. Chengalur, S.A. Pustilnik]
{J.N. Chengalur,$^{1}$\thanks{E-mail: chengalur@ncra.tifr.in (JNC), sap@sao.ru (SAP)}
S.A. Pustilnik$^{2,3}$ \\
\rule{-4pt}{20pt}
$^1$ National Centre for Radio Astrophysics, Post Bag 3, Ganeshkhind, Pune
  411 007, India\\
$^2$ Special Astrophysical Observatory of RAS, Nizhnij Arkhyz,
  Karachai-Circassia 369167, Russia\\
$^3$ Isaac Newton Institute of Chile, SAO branch, Nizhnij Arkhyz, Russia}

\begin{document}

\label{firstpage}

\date{Accepted 2012 September ??. Received 2012 August 23}

\pagerange{\pageref{firstpage}--\pageref{lastpage}} \pubyear{2012}

\maketitle

\begin{abstract}

Giant Metrewave Radio Telescope (GMRT) \HI\ observations, done as part
of an ongoing study of dwarf galaxies in the Lynx-Cancer void, resulted
in the discovery of a triplet of extremely gas rich galaxies located
near the centre of the void.The triplet members SDSS J0723+3621, J0723+3622
and J0723+3624 have absolute magnitudes $M_{\rm B}$ of --14.2, --11.9 and
--9.7 and $M$(\HI)/$L_{\rm B}$ of $\sim$2.9, $\sim$10 and $\sim$25,
respectively. The gas mass fractions, as derived from the SDSS photometry
and the GMRT data are 0.93, 0.997, 0.997 respectively. The faintest member 
of this triplet SDSS~J0723+3624 is one of the most gas rich galaxies
known. We find that all three galaxies deviate significantly from the
Tully-Fisher relation, but follow the baryonic Tully-Fisher relation.
All three galaxies also have a baryon fraction that is significantly 
smaller than the cosmic baryon fraction. For the largest galaxy in the triplet,
this is in contradiction to numerical simulations. The discovery of
this very unique dwarf triplet lends support to the idea that
the void environment is conducive to the formation of galaxies with
unusual properties. These observations also provide further motivation to
do deep searches of voids for a ``hidden'' very gas-rich
galaxy population with $\mb \gtrsim -11$.

\end{abstract}


\section[]{Introduction}
\label{sec:intro}

Early redshift surveys established that the spatial distribution of
bright galaxies is highly inhomogeneous and that the properties of the
galaxy population varies with environment.  The spatial distribution 
was found to consist of large under dense regions (``voids'') surrounded
by galaxies in sheets and walls \citep{joeveer78,kirshner81, geller89}.
Further,
the morphological mix of galaxies was found to vary systematically
with galaxy density. The fraction of late type galaxies
monotonically increases as one goes from high density to low density
regions \citep{postman84}. Subsequent numerical simulations showed 
that the existence of voids can be understood as a consequence of 
biasing in the formation of galaxy halos, with the most massive halos 
being formed in regions of high densities \citep{white87}. However,
numerical simulations also predicted that the voids should be 
filled with small mass halos \citep[e.g.][]{davis85,gottlober03}. 
\citet{peebles01} pointed out that, contrary to this expectation, 
the known dwarf galaxies instead follow the same large scale
distribution as the bright galaxies, a discrepancy that he dubbed
the ``void phenomenon''. We note that earlier \citep{pustilnik95}
as well as recent \citep{kreckel12} studies do find dwarf galaxies
inside voids, however it remains true that the brighter dwarfs generally
lie near the void walls. \cite{peebles01} also highlighted that if
the small dark halos produced in voids preferentially fail to 
host galaxies, this would correspond to a discontinuous change in 
galaxy properties with density. This distinguishes the ``void phenomenon''
from the observed ``morphology-density'' relationship in which the
morphological mix varies smoothly with density. Observational and 
theoretical studies of the void galaxy population have since been largely 
focused on these two (possibly related) issues viz. (i)~a search for 
the ``missing'' dwarfs in voids and (ii)~the influence of the large 
scale environment on galaxy properties.

Regarding the issue of missing dwarfs, deep searches of voids have shown
that they do not contain dwarf galaxies in the numbers predicted by
simulations \citep[see e.g.][]{tikhonov09,kreckel12}. The reason
for this discrepancy is unclear, although there have been numerous
suggestions that the formation of galaxies in small dark matter halos
in voids is suppressed (e.g. \citet{tinker09}; see also \citet{kreckel11}).
Essentially, if small halos in voids are baryon deficient this would
resolve the problem of missing dwarfs. \citet{hoeft10} examined the 
baryon fraction of small halos in both voids and filaments using high
resolution simulations, and found no dependence of the baryon deficiency
on environment. The discrepancy between numerical simulations and 
the observations hence remains a puzzle. 

Regarding the issue of the effect of the large scale environment on 
the properties of void galaxies, studies using SDSS selected samples 
established that the void galaxy population is significantly bluer 
and has a higher star formation rate as compared to the high density
galaxy population \citep{rojas04,rojas05}. However \citet{patiri06}
(see also \citep{park07}) show that this difference is almost entirely 
due to the morphology-density relation. Late type galaxies which are more 
dominant in low density regions, have bluer colours and higher star 
formation rates than early type galaxies. At a fixed luminosity and 
morphology the properties of the detected void galaxies are statistically
identical to that of galaxies in dense regions. It is worth noting here
that these conclusions relate only to the upper part of the whole luminosity
(or mass) range of void galaxies ($M_{\rm B,r} \lesssim -16$~mag) and
do not include study of possible differences in parameters such as the
gas phase metallicity and gas mass fraction.

The gas mass fraction of void galaxies was studied in earlier works which 
looked at the distribution of $M$(\HI)/$L_{\rm B}$. \citet{huchtmeier97} 
found that dwarf galaxies closer to the center of the void had a higher
$M$(\HI)/$L_{\rm B}$ than galaxies close to the walls.  Similarly, 
\citet{pustilnik02} found marginal evidence for low luminosity 
galaxies in voids to have a higher $M$(\HI)/$L_{\rm B}$ than those in
higher density regions. Extrapolation of the obtained trends to
the range $M_{\rm B} > -15$ indicated that for lower mass dwarfs the
difference could be higher. On the other hand, \citet{kreckel12} show
that the \HI\ gas content of void galaxies are statistically 
indistinguishable from galaxies in filaments and walls, at least
for galaxies brighter  than $M_{\rm r} \sim -16$~mag. This result does not
contradict earlier results, and underlines the need of deeper void
galaxy samples. 

A possible resolution of the discrepancy between the predictions of
the numerical simulations and the observations is that the dwarfs predicted
to exist in voids are fainter than what the observations have probed so
far. For example, in their simulation, \citet{kreckel11} find that while 
luminous dwarfs (\mr\ brighter than $\sim -18$~mag) in voids are statistically 
indistinguishable from similar dwarfs in higher density regions,
fainter dwarfs (\mr\ $\sim -16$~mag) are significantly bluer and have higher
specific star formation rates than their higher density counter parts. 
They also find a significant excess of faint dwarf (\mr $\sim -14$~mag) galaxies
that are preferentially located in low density regions near the void centre. 
To complement their numerical simulations \citet{kreckel12} used the SDSS
to identify a population of void galaxies with $\mr > -16.1$~mag. To identify 
and study still fainter objects one needs to focus on the nearby
voids.

In a recent series of papers (\citet{pustilnik11a}~(Paper~I),
\citet{pustilnik11b} (Paper~II), \citet{pustilnik11c}~(Paper~III)) 
a sample of 79 galaxies residing in the nearby Lynx-Cancer void
was presented. The sample galaxies have \mb\  down to -12~mag, with 
reasonable completeness level at $\mb \sim -14.0$~mag. For this 
completeness level, the average void galaxy density ($\sim$0.04~Mpc$^{-3}$) 
is about one order of magnitude smaller than the mean value for the faint SDSS 
galaxies derived by \citet{blanton05}. More than half of the Lynx-Cancer void 
sample consists of low surface brightness dwarfs (LSBDs). Measurements of
O/H are available for $\sim 60$\% of the sample, and shows that the 
metallicity of the void galaxies is on average $\sim 30\%$ lower
than that of their counterparts in denser regions. About 10\% of the sample
are deficient in metals by factors of 2-7. A GMRT based \HI\ study of 
these dwarfs is in progress. Here we report on a highly unusual system 
found in the course of the \HI\ observations, viz. an extremely gas rich 
triplet of LSBD galaxies, located inside the central 10\% of the void 
volume.

\section[]{Observations and Results}
\label{sec:obs}

\begin{table}
\caption{Parameters of the GMRT observations}
\label{tab:obspar}
\begin{tabular}{ll}
\hline
     & J0723+36 triplet  \\
\hline
Date of observations     & 2011 Nov 25  \\
Field center R.A.(2000)  &07$^{h}$23$^{m}$07.40$^{s}$   \\
Field center Dec.(2000)  &+36$^{o}$22$^{'}$41.0$^{"}$    \\
Central Velocity (\kms)  & 950.0   \\
Time on-source  (h)      &$\sim$6  \\
Number of channels       & 256  \\
Channel separation (\kms)& $\sim$1.73 \\
Flux Calibrators         & 3C48,3C286 \\
Phase Calibrators        & 0741+312  \\
Resolution (arcsec$^{2}$)& 40~$\times$~40 \\
rms (mJy~Bm$^{-1}$)      & 2.8\\
\hline
\hline
\end{tabular}
\end {table}

GMRT \HI\ 21cm observations of the J0723+36 system were conducted on
25 Nov. 2011. The observational parameters are given in
Table~\ref{tab:obspar}. The initial flagging and calibration was
done using the flagcal package \citep{prasad12} and further
processing was done using standard tasks in the AIPS package. 
The hybrid resolution of the GMRT allows one to make maps at 
various resolutions, however, here we show only maps at 
$40^{''}$ resolution.

\begin{figure}
 \includegraphics[angle=-90,width=8.0cm]{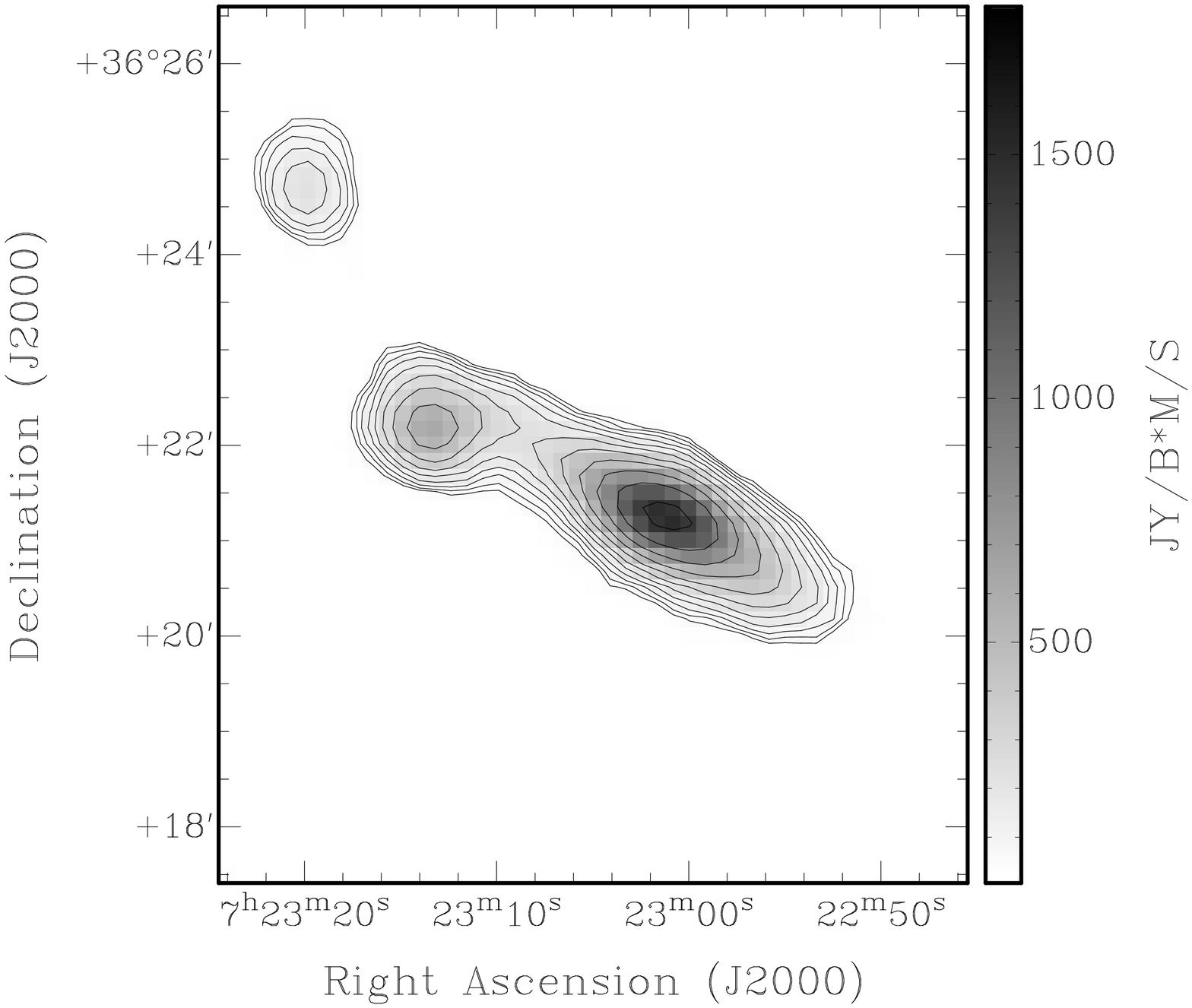}
 \includegraphics[angle=-90,width=8.0cm]{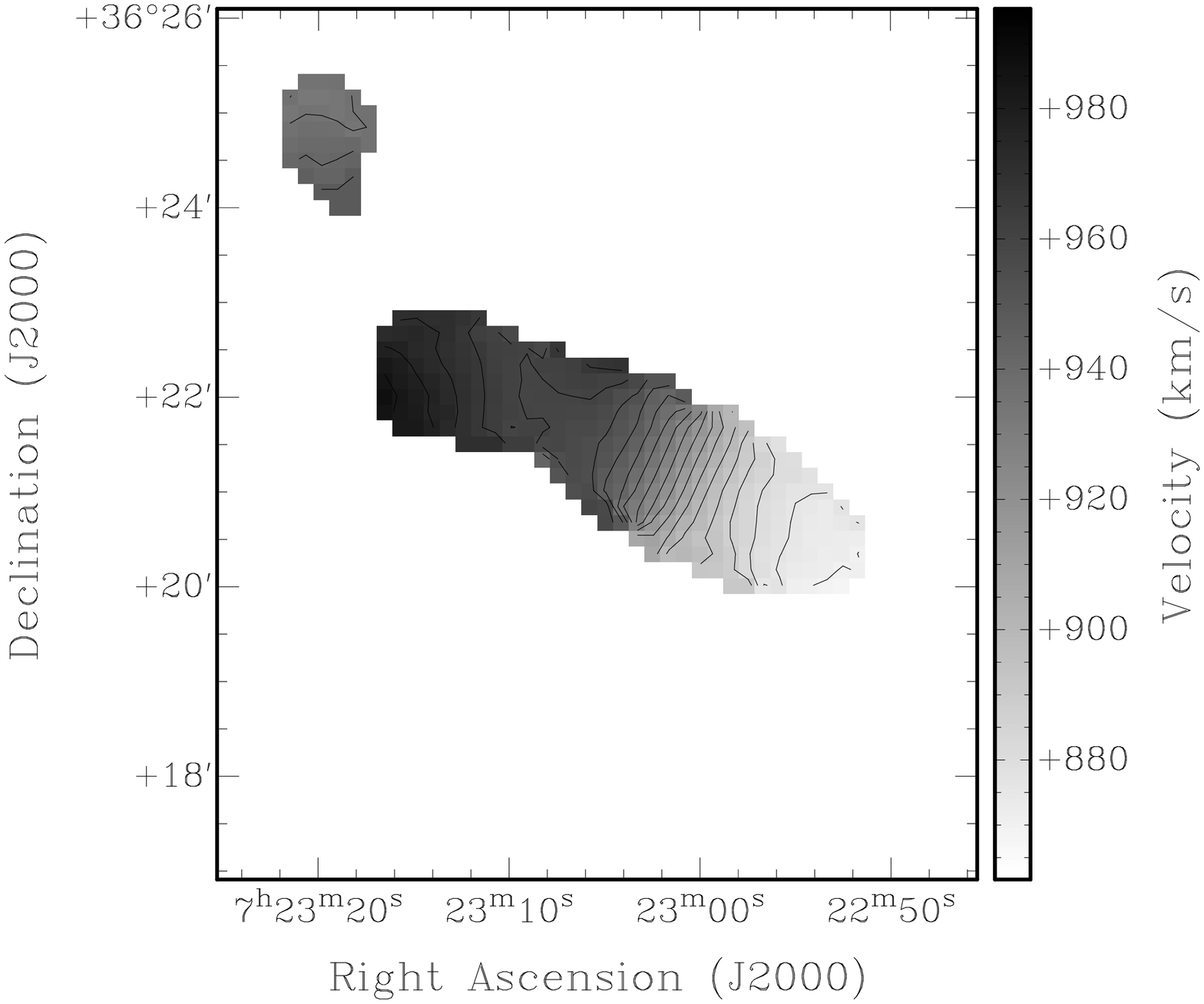}
 \caption{\label{fig:HImaps}
[A] The integrated \HI\ emission (moment0 map) from the J0723+36 system.
    The angular resolution is 40$^{''}$. The contours start at
    $3\times 10^{19}$ atoms~cm$^{-2}$ and are in steps of 1.414.
[B] The velocity field (moment1 map) of the J0723+36 system, derived 
    from the 40$^{''}$ resolution data. The iso-velocity contours
    start at 870.0~\kms\ and are in steps of 6~\kms.}
\end{figure}

\begin{figure}
  \includegraphics[angle=-90,width=7.0cm]{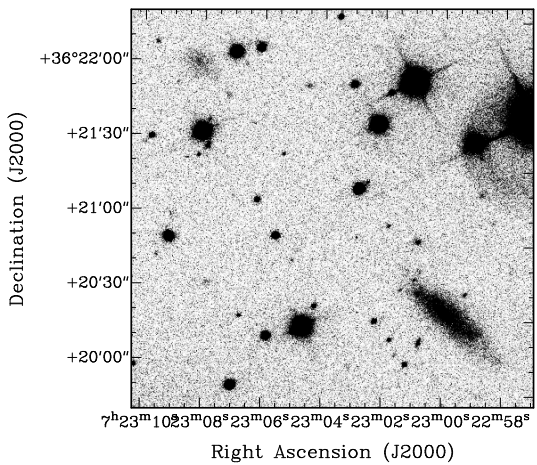}
  \caption{\label{fig:SDSS}
	  The SDSS $g$ band image showing the main two galaxies in the
	 triplet, viz. J0723+3621 (the edge on galaxy) and J0723+3622
         (the fainter companion).}
\end{figure}

\begin{figure}
  \includegraphics[angle=-90,width=7.0cm]{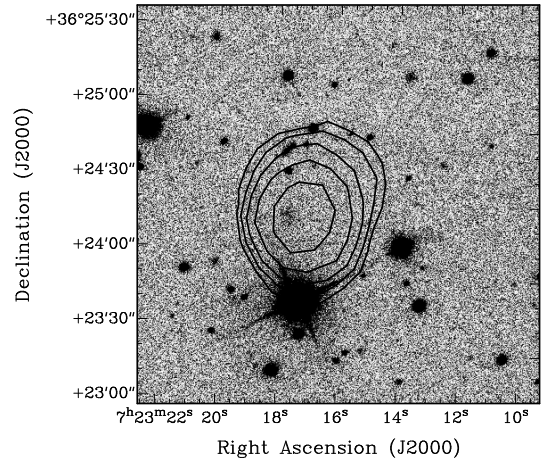}
  \caption{\label{fig:SDSScomp}
	  The SDSS $g$ band image showing  the faintest member of the triplet,
          J0723+3624. The GMRT \HI\ contours are also overlayed, the contour
          levels are the same as in Fig.~\ref{fig:HImaps}.}
\end{figure}

In Fig.~\ref{fig:HImaps}[A] is shown the GMRT \HI\ map of the J0723+36 
system. At the time of the observations, only two galaxies, viz.
SDSS~J0723+3621, and SDSS~J0723+3622 were known to lie within the observed
data cube. At the adopted distance of 16~Mpc (see below) to this group  the 
separation of the pair is 12.1~kpc. As can be seen from the figure, the 
GMRT observations detected one more \HI\ source, which corresponds to 
the galaxy SDSS~J0723+3624. The projected separation between this 
companion and the brighter of the two galaxies in the pair (viz.
J0723+3621) is 23.9~kpc. SDSS $g$ band images showing these
galaxies is shown in Fig.~\ref{fig:SDSS} and Fig.~\ref{fig:SDSScomp}
(a colour composite for this pair is also shown in Paper~III).
The two brighter galaxies, J0723+3621, and J0723+3622 are
clearly interacting, with a bridge of \HI\ connecting them. For the  
third much smaller system there is a hint of an extension to the North-West,
but the resolution is marginal. The velocity field (with isovelocity lines 
in steps of 
6.0~\kms) for the whole system is shown in Fig.~\ref{fig:HImaps}[B]. All 
three galaxies show clear velocity gradients, although in all cases 
the velocity field is disturbed. In the case of the galaxy pair, 
this is clearly due to the ongoing tidal interaction. The spins of both
components of the pair are aligned with their orbital angular momentum, 
consistent with the pair undergoing a prograde collision. A prograde 
encounter geometry is also consistent with the significant tidal 
distortions seen. 

A continuum image made using all the available channels shows no emission 
from the triplet galaxies. The rms noise level is $\sim 0.5$~mJy/Bm at
an angular resolution of $40^{''}$.

\begin{figure*}
 \centering
 \includegraphics[angle=0,width=5.80cm]{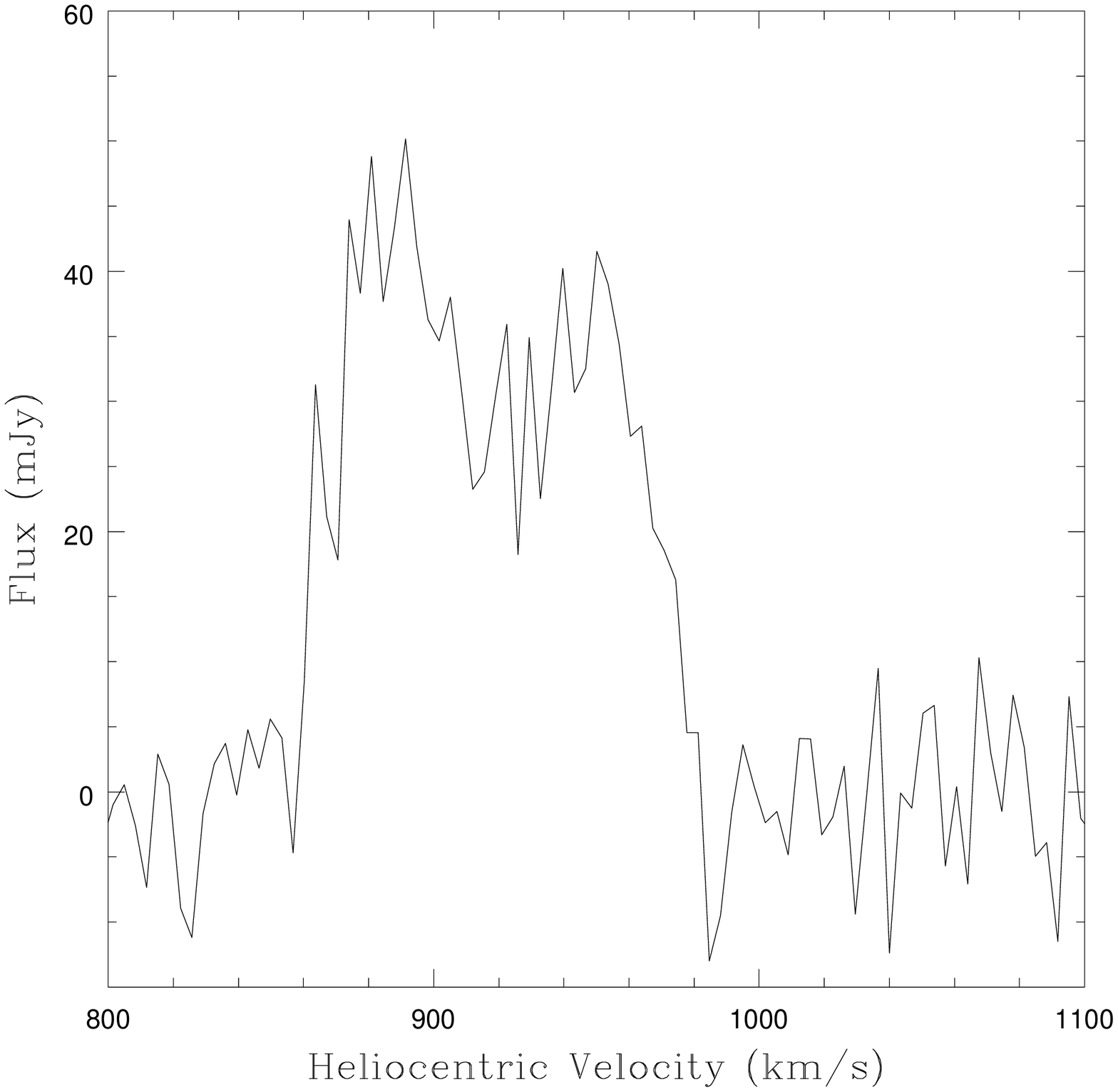}
 \includegraphics[angle=0,width=5.80cm]{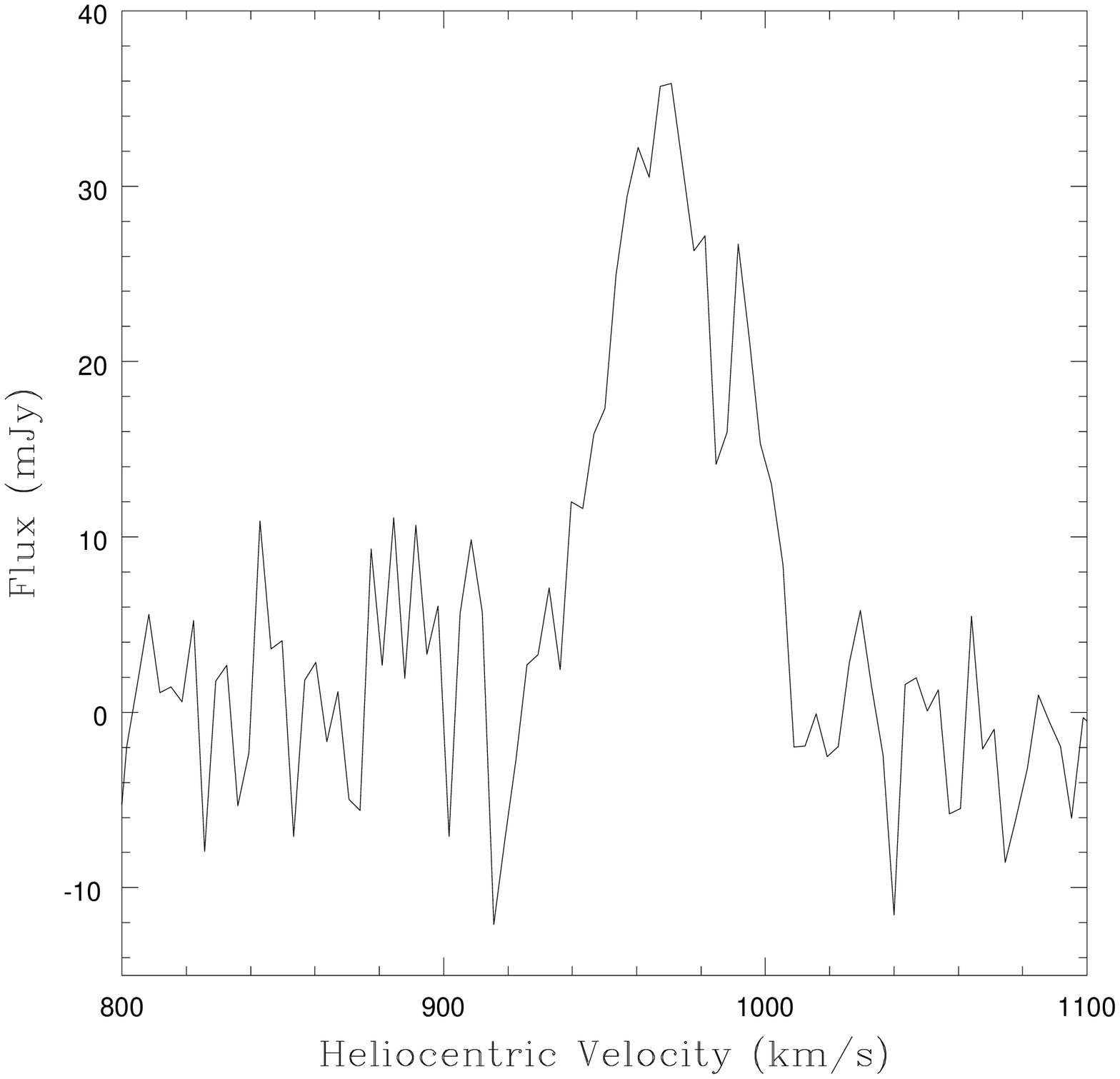}
 \includegraphics[angle=0,width=5.80cm]{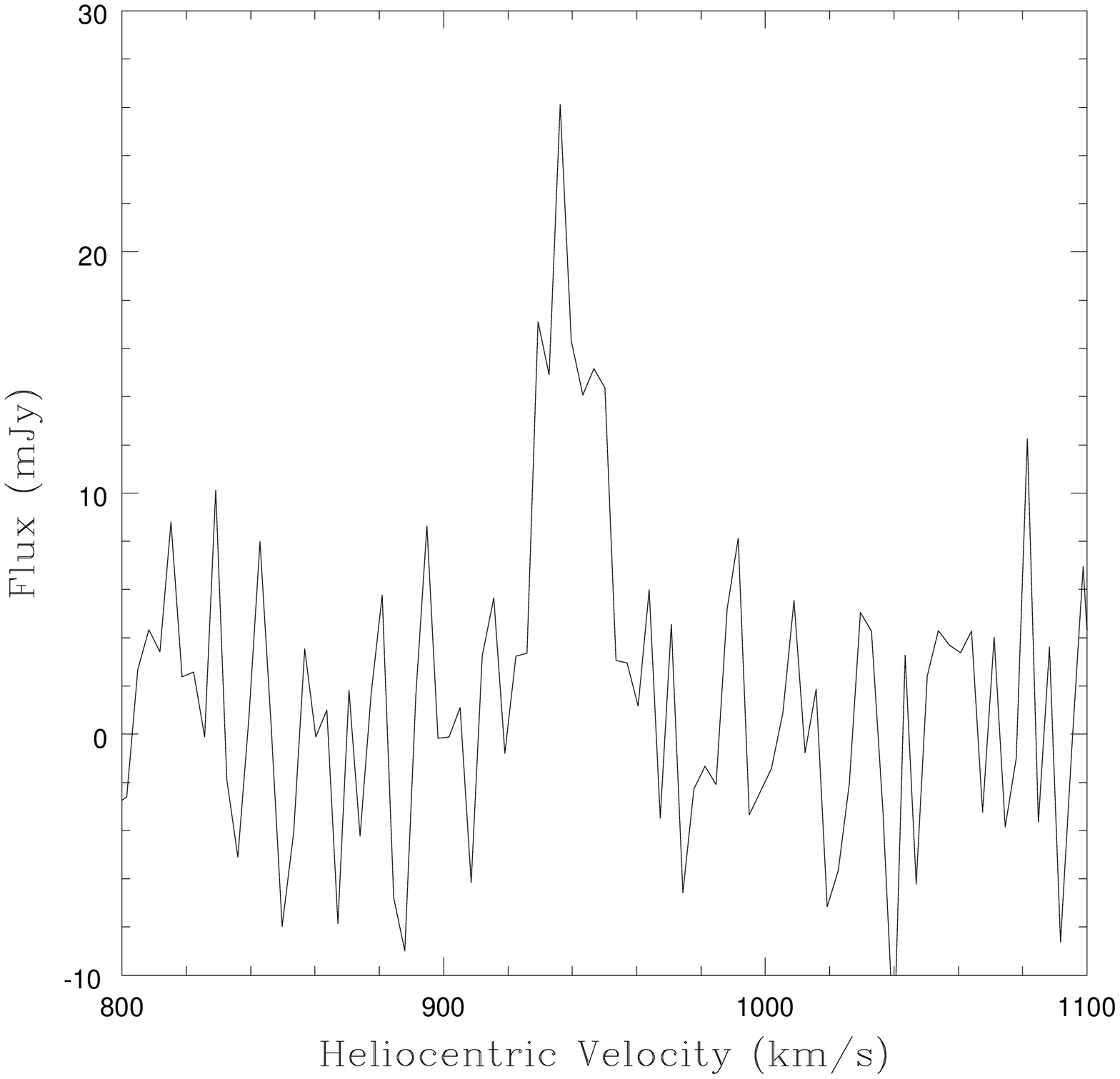}
  \caption{\label{fig:HIprofiles}
Integrated \HI\ spectra for the members of the J0723+36 triplet.
The spectra have been derived from the 40$^{''}$ resolution data
cube. The left panel is for J0723+3621. Note that the first profile
has a double horned shape, characteristic differentially rotating
disks. According to Geha et al. (2006), 18\% of dwarfs with 
$M_{\rm r} > -16$ have such profiles. The middle panel is for
J0723+3622 while the right panel is for J0723+3624.
}
\end{figure*}

The main parameters of the three galaxies in the 0723+36 system are
given in Tab.~\ref{tab:param}. The \HI\ parameters are derived from the
integrated single dish profiles shown in Fig~\ref{fig:HIprofiles}. 
Because the velocity fields are disturbed, we do not attempt to 
derive rotation curves. Instead we use the velocity widths obtained
from the integrated profiles to estimate the dynamical mass related 
quantities. The optical parameters are derived from the SDSS DR7 data
\citep{DR7}.
In the case of the newly discovered companion galaxy J0723+3624, 
there are two extended SDSS objects, separated by $\sim$5.5\arcsec\ 
(0.4 kpc in projection) seen near the center of the \HI\ emission. 
We refer below to these two objects associated with the companion 
J0723+3624 as the NE and SW components. The NE component 
(J072320.57+362440.8), has $g$=21.42 and somewhat blue colours, 
($(g-r)_{\mathrm 0} =0.09\pm0.19$), albeit with large error bars. 
Independent photometry of this object based on the SDSS images 
resulted in very similar values, viz. $g$=21.29$\pm$0.05 and 
$(g-r)_{\mathrm 0} =0.08\pm0.08$. The SW component J072320.32+362436.7 
is $\sim$1.3~mag fainter in $g$-filter and is significantly redder
($(g-r)_{\mathrm 0} =0.63\pm0.28$). Independent measurements from the
SDSS data once again result in similar values, viz.  a $g$-flux that
is $\sim$ 1.2-mag fainter and $(g-r)_{\mathrm 0} =0.60\pm0.11$.
This `red nebulosity' looks similar  to the very faint red galaxies seen
to the North and East of the blue component, which could represent a small
group of distant galaxies. We hence assume that the red nebulosity 
is unrelated to the \HI-cloud. In any case, including this component 
would make only a minor difference to the total flux -- making
the object brighter in the $B$-band by 0.26~mag. 

The rows of Tab.~\ref{tab:param} are as follows:
the J2000 RA and Dec;
$A_{\rm B}$, the  Galactic extinction in $B$-band; 
the total $B$-magnitude, not corrected for $A_{\rm B}$, obtained by 
transformation from the total $g$ and $r$, according to the formulae
given in Lupton et al. (2006);
the $(g-r)$ colour;
the $(B-V)$ colour, computed from the observed $(g-r)$ and for the 
PEGASE2 constant evolutionary tracks with $z=0.002$ (or $Z$ = $Z$\sunn/10).
the heliocentric velocity, obtained from \HI-profile in this paper; 
the adopted distance, accounting for the updated $V_{\rm hel}$ and the 
large negative velocity correction, described in Paper~I;
the calculated absolute blue magnitude, optical sizes (angular and linear), 
corrected for the Galactic extinction and inclination;
the central inclination corrected surface brightness in $B$-band;
the measured \HI-flux in units of Jy~\kms;
the profile widths $W_{\mathrm 50}$ and $W_{\mathrm 20}$;
the inclination angle estimated from the SDSS images;
the \HI\ mass $M$(\HI); the total baryonic mass computed as
$ M_{\rm bary} = 1.3 \times M(\HI) + M_{\rm star}$.
$M_{\rm star}$ is the stellar mass computed from the SDSS data using the
total magnitude in $g$-filter and the total $(g-i)$ colour, corrected 
for the Galactic extinction (from NED, following to Schlegel et al. 1998),
and the mass to luminosity ratio $\Upsilon$ determined from the 
prescription given in Zibetti et al. (2009) (similar to used in Paper~III);
the total dynamical mass, computed using the formula 
M$_{\rm dyn} = 2.3 \times 10^5 \times R_{\rm kpc} \times V_{\kms}^2$,
where $R_{\rm kpc}$ is the radius measured from the \HI\ images at the
$3 \times 10^{19}$~atoms~cm$^{-2}$ level, and $V_{\kms}$ is the 
rotational velocity computed from $W_{\mathrm 20}$ after correction for
inclination and turbulent motions using the prescription given 
in \citet{verheijen01};
$M_{\rm vir}$, the virial mass, computed from the circular velocity estimated
as above, and the formulae given in \citet{hoeft06};
$R_{\rm vir}$, the virial radius, computed from the circular velocity
estimated
as above, and the formulae given in \citet{hoeft06};
the ratio of \HI\ mass to blue luminosity, $M$(\HI)/$L_{\rm B}$ in
solar units;
the gas mass fraction $f_{\rm gas} = 1.3 \times M(\HI)/M_{\rm bary}$;
the baryon fraction $f_{\rm bar} = M_{\rm bar}/M_{\rm vir}$.

\begin{table}
\caption{Main parameters of the J0723+36 triplet}
\label{tab:param}
\begin{tabular}{lccc} \\ \hline
Parameter                           &  J0723+3621             & J0723+3622       & J0723+3624     \\ \hline
R.A.(J2000.0)                       & 07 23 01.42             & 07 23 13.46      & 07 23 20.57    \\
DEC.(J2000.0)                       & $+$36 21 17.1           & $+$36 22 13.0    & $+$36 24 40.8  \\
$A_{\rm B}$ (from NED)               & 0.23                    & 0.23             & 0.23           \\
$B_{\rm tot}$                        & 17.01$\pm$0.03          & 19.31$\pm$0.03   & 21.56           \\
$(g-r)_{\rm 0,tot}$                 & 0.34$\pm$0.01            & --0.12$\pm$0.11   & 0.08$\pm$0.12\\
$(B-V)_{\rm 0,tot}$                 & 0.34$\pm$0.01            & 0.02$\pm$0.08    & 0.17$\pm$0.09 \\
$V_{\rm hel}$(\HI)(\kms)             & 917$\pm$1               & 970$\pm$1        & 938$\pm$1       \\
Distance (Mpc)                      & 16.0                    & 16.0             & 16.0            \\
$M_{\rm B}^0$                        &  --14.24                & --11.94          & --9.68         \\
Opt. size (\arcsec)$^{5}$           & 44$\times$15$^{(2)}$    & 12$\times$9.5    & 6$\times$4:     \\
Opt. size (kpc)                     & 3.41$\times$1.16$^{(2)}$& 0.93$\times$0.74 &0.47$\times$0.31 \\
$\mu_{\rm B}^0$(mag~arcsec$^{-2}$)   & 24.14                   & 24.36            &  24.6:         \\
\HI\ int.flux                       & 3.74$\pm$0.4            & 1.59$\pm$0.2     & 0.48$\pm$0.05   \\
$W_\mathrm{50}$ (km s$^{-1}$)        & 100.2$\pm$0.7           & 45.3$\pm$0.7     & 22.0$\pm$0.7    \\
$W_\mathrm{20}$ (km s$^{-1}$)        & 122.5$\pm$1.0           & 69.0$\pm$1.0     & 33.5$\pm$1.0    \\
$i$ (degrees)                       & 90:                     & 60:              & 60:             \\
$V_{\rm rot}$ (\HI)(\kms)         & 54.0                    & 35.2             & 15.3            \\
$M$(\HI) (10$^{7}~M_{\odot}$)        & 22.6                    & 9.61             & 2.9             \\
$M_{\rm bary}$ (10$^{7}~M_{\odot}$)  & 32.37                   & 12.8             & 3.86            \\
$M_{\rm dyn}$ (10$^{7} M_{\odot}$)   & 623.8                   & 132.5            & 16.91           \\
$M_{\rm vir}$ (10$^{8} M_{\odot}$)   & 359.5                   & 102.3            & 8.9           \\
$R_{\rm vir}$ (kpc)                  & 87.3                    & 57.4             & 25.4            \\
$M$(\HI)/$L_{\rm B}$                 & 2.9                     & 10.2             & 25              \\
$f_{\rm gas}$                        & 0.93                    & 0.997            & 0.997           \\
$f_{\rm bar}$                        & 0.009                   & 0.013           & 0.043          \\
\hline
\end{tabular}
\end{table}

\section[]{Discussion}
\label{sec:dis}

\begin{figure}
  \includegraphics[angle=-0,width=8.0cm]{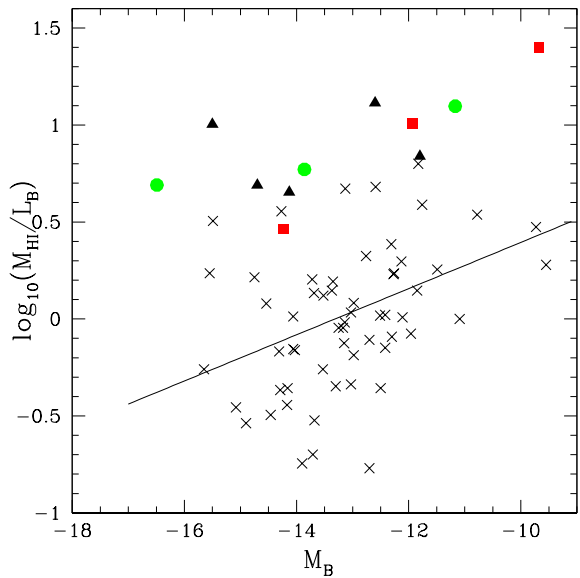}
  \caption{\label{fig:mhilb} Comparison of the $M$(\HI)/$L_{\rm B}$ ratio of
the galaxies in the Lynx-Cancer triplet (filled squares), with
galaxies from the FIGGS sample \citep[][crosses]{begum08}. Three very gas
rich dwarfs from the Void Galaxy Sample (VGS) of \citet{kreckel12}
are shown as  circles, and data for 6 other extremely gas rich
galaxies (see Tab.~\ref{tab:mostgasrich}) are shown in triangles.
The solid line is the best fit relation for the FIGGS galaxies.
}
\end{figure}

As can be seen from Tab.~\ref{tab:param}, all three galaxies have very
large $M$(\HI)/$L_{\rm B}$ ratios; in fact J0723+3624 has one of the 
larges ratios
known. The corresponding gas mass fractions are also 
extremely large. Even if we assume that the colours of J0723+3622 and 
J0723+3624 are redder by $1\sigma$ that the measured values, their gas 
mass fractions remain $\gtrsim$0.99. For star forming galaxies the 
gas mass fraction is known to
increase with decreasing luminosity \citep[see e.g.][]{mcgaugh97,geha06}. 
We show in Fig.~\ref{fig:mhilb} $M$(\HI)/$L_{\rm B}$ for the FIGGS 
(\citep{begum08}) sample, which also clearly shows this trend. The data for
the galaxies from the J0723+36 triplet are also shown, and, as can 
be seen, for their given luminosities, all three galaxies lie at the
extreme gas rich end of the distribution.

\begin{figure*}
 \centering
 \includegraphics[angle=-0,width=8.5cm]{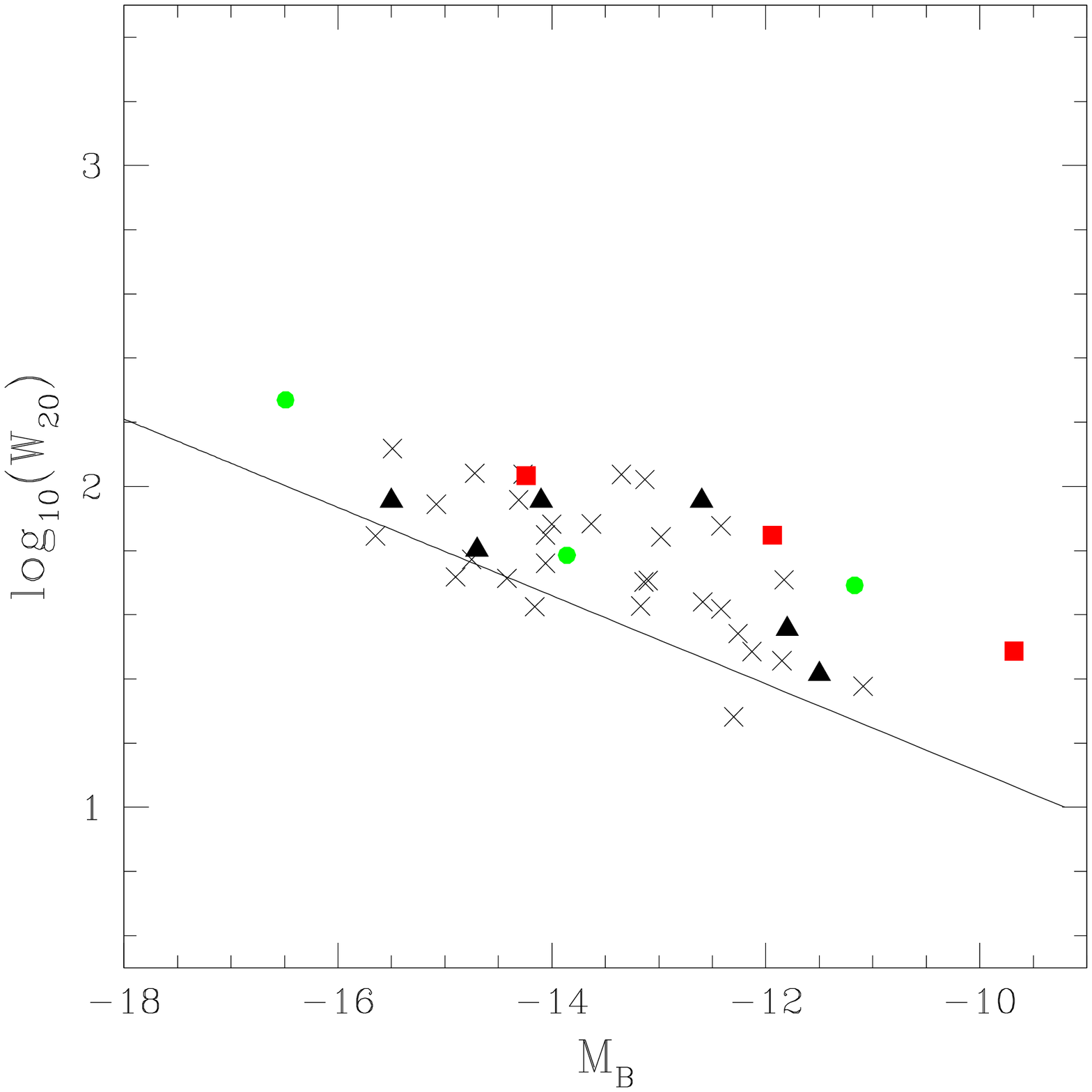}
 \includegraphics[angle=-0,width=8.5cm]{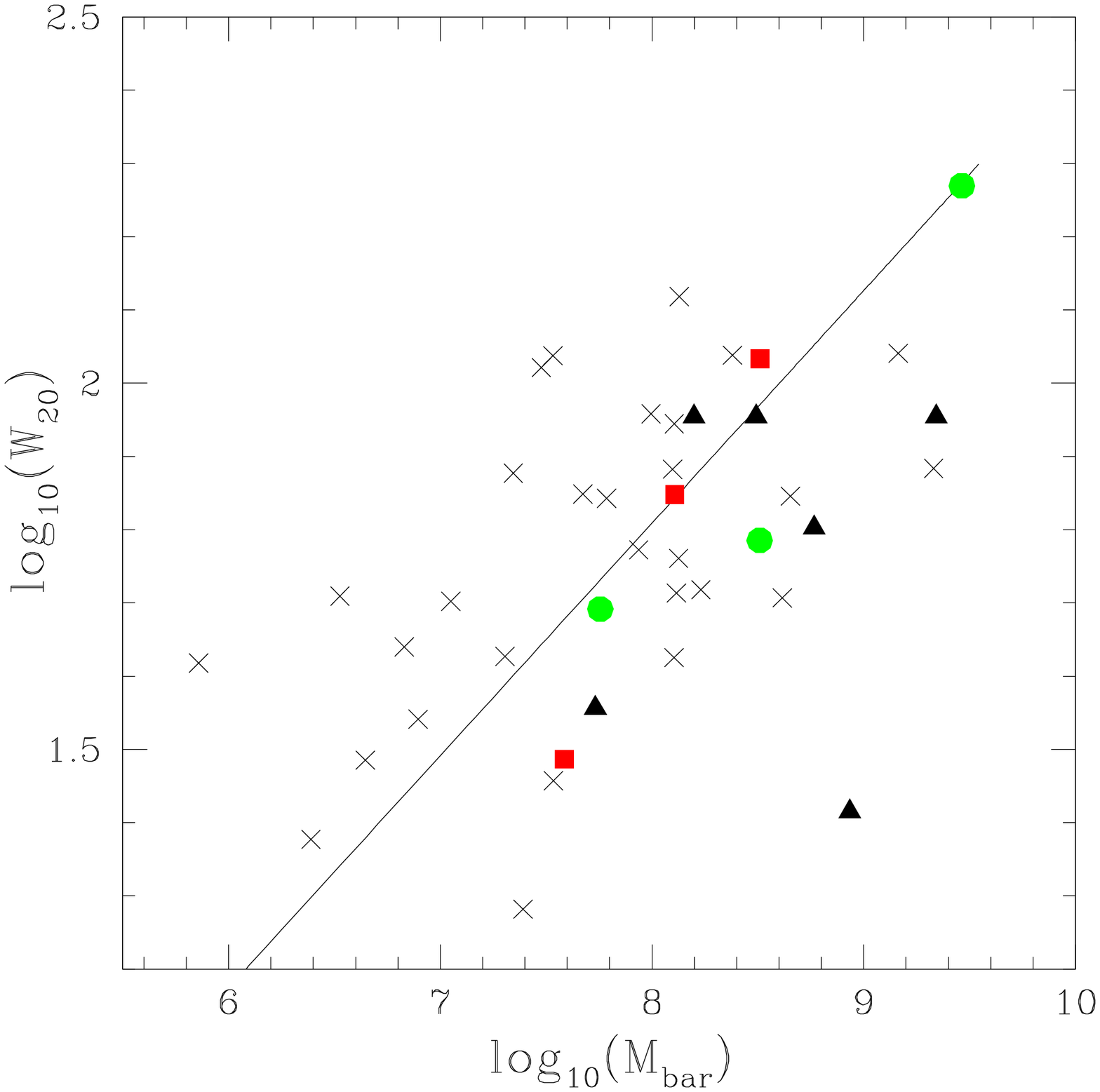}
  \caption{\label{fig:tf} 
{\bf [A]}:  The Tully-Fisher relation: $W_{\mathrm 20}$ vs \mb.
The reference TF relation line (solid) for brighter galaxies \citep{tully00}.
{\bf [B]:} The Baryonic Tully-Fisher relation. The solid line shows the 
BTF relation from \citet{derijcke07}.
The symbols are the same as those used in Fig.~\ref{fig:mhilb}
}
\end{figure*}

We also compare the galaxies location in the Tully-Fisher (TF) and Baryonic
Tully-Fisher (BTF) diagrams. We note that the inclinations estimated for these
galaxies are somewhat uncertain, however this should affect the TF and
BTF relations equally. In Fig.~\ref{fig:tf}[A] is shown the Tully Fisher
relation for the FIGGS galaxies. Also overplotted is the TF relation 
for bright galaxies, as determined by \cite{tully00}. As expected for 
dwarf galaxies, the FIGGS galaxies are underluminous for their velocity 
width \citep[ see also][]{begum08a}. Once again, the triplet galaxies
fall at the extreme end of the 
distribution. Fig.~\ref{fig:tf}[B] shows the baryonic Tully-Fisher relation, 
with for reference the BTF relation shown from \citet{derijcke07}. 
Despite being  extremely gas rich, the triplet galaxies do follow 
the BTF relation.

The baryon fraction of small galaxies is also an interesting quantity to
compare against model predictions. For our extremely gas rich galaxies,
the baryon fraction can be accurately measured. From numerical models 
\citep{hoeft06,hoeft10}, one would 
expect that galaxies with circular velocities $\gtrsim$~50~\kms\ would
have a baryon fraction equal to that of the cosmic value of $\sim 0.17$
\citep{jarosik11}. For smaller galaxies, the baryon fraction drops
sharply with circular velocity, at a circular velocity of 20~\kms, the
predicted baryon fraction is an order of magnitude below the cosmic
value. As can be seen from Tab.~\ref{tab:param}, for all galaxies
the baryon fraction $f_{\rm bar}$ is more than an order of magnitude
lower than the cosmic baryon fraction.  In the case of the brightest
galaxy J0723+3621, the baryon fraction is $\sim 1/20$ that of the
cosmic baryon fraction, while one would expect it to have the 
cosmic baryon fraction. This galaxy is clearly edge on, and 
this result is hence unlikely to be due to an uncertain inclination
angle. \citet{mcgaugh10} has earlier highlighted
that the measured baryon fraction for galaxies with rotation
velocities in this range is significantly smaller than the cosmic
baryon fraction. In this respect, J0723+3621 is similar to dwarf
galaxies located outside of voids. Interestingly, $f_{\rm bar}$ appears
to decrease with increasing velocity width, i.e. the reverse of
what is predicted. One possible reason for this could be that
for the smaller galaxies the baryons do not sample the flat part
of the rotation curve, and hence $W_{\mathrm 20}$ underestimates the
circular velocity. It is also worth noting that the expected
virial radius of even the smallest galaxies dark matter halo
is larger than the separation of the
galaxies in this triplet. If one regards the entire triplet as
a single system then the  $f_{\rm bar}$ is $\sim 0.01$, about
16 times smaller than the cosmic baryon fraction.

The triplet of galaxies that we are discussing here are found in the
inner 10\% volume of the Lynx-Cancer void. \citet{kreckel12} also
find 3 similarly gas rich galaxies (VGS\_7a, VGS\_9a, VGS\_12) in
their void survey. Their absolute blue magnitudes computed from
the SDSS magnitudes following the same procedures as for our triplet
galaxies are \mb\ of --13.86, --11.17 and -16.49 respectively.
The corresponding $M$(\HI)/$L_{\rm B}$ ratios are is $\sim$5.9, $\sim$12.5
and $\sim$4.9. These  very gas rich galaxies found in 
surveys of voids are interesting in view of the possibility highlighted
by  \citet{peebles01} that the void environment could be conducive 
to the production of galaxies with extreme properties. 

To examine this issue further, we take a look at the eleven galaxies 
with well measured $M$(\HI)/$L_{\rm B} \gtrsim$ 5 that we are aware of.
The sample
consists of three galaxies from the \citet{kreckel12} VGS sample, the 
two fainter members of the triplet discussed in this paper and six 
galaxies taken from the literature (see Tab.~\ref{tab:mostgasrich}). 
For the two members of Lynx-Cancer void triplet and three most gas-rich VGS
galaxies the type of global environment is clear from the sample
selection. For the 6 galaxies from Tab.~\ref{tab:mostgasrich} the
environment varies. The nearest two objects, UGC~292 and DDO~154, 
are situated on the periphery of the loose aggregate named Canis Venatici~I 
Cloud (Karachentsev et al. 2003), which probably is still in 
the formation phase. The next most distant galaxy And~IV is situated far 
from luminous galaxies, at $D \sim$6.1~Mpc, and according to
\citet{Ferguson00} probably belongs to a loose dwarf group. However, its 
projected distances to the brightest dwarfs of this group are of more
than 2~Mpc.
The distance to the unusual 
pair of very gas-rich objects HI~1225+01~SW,NE is very uncertain, since
it is unclear if this object lies in the foreground or background
of the Virgo cluster. Because of this distance uncertainty, it is not
possible to make a definitive statement about its environment. However, after 
a detailed analysis \citet{salzer92} conclude that there is no 
evidence for a massive neighbour within $\sim 1$~Mpc of the system.
The SBS~0335--052 system is at a projected separation of only 150~kpc from
the large spiral galaxy NGC~1376 \citep{pustilnik01}. \citet{peebles01} 
examined its location with respect to galaxies from the ORS 
\citep{santiago95}, and concluded that while it is not located in a 
dense region, neither is it located in a void. In summary, while all 
of the gas rich galaxies appear relatively isolated, it is not the 
case that they are {\it all} located deep inside voids. 

\begin{figure}
 \centering
 \includegraphics[angle=-90,width=9.0cm]{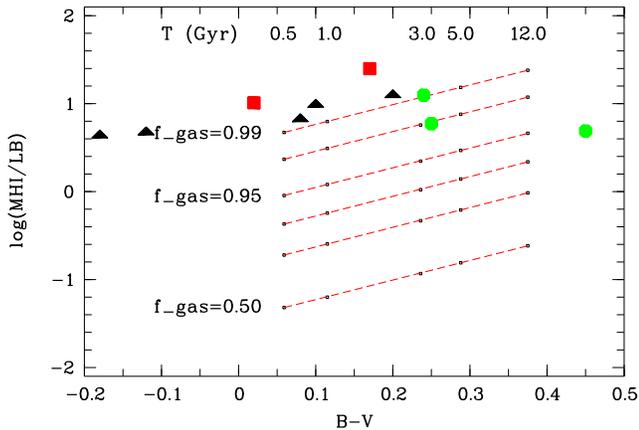}
  \caption{\label{fig:pegase} 
Constant star formation rate PEGASE2 models (dashed) of $\log$($M$(\HI)/$L_{\rm B}$)
versus ($B-V$)  for galaxies with constant gas mass-fraction $f_{\rm gas}$.
Model lines are shown for $f_{\rm gas}$ of 0.50, 0.80, 0.90, 0.95, 0.98 and 0.99.
The lines connect points corresponding to ages of 0.5, 1.0, 3.0, 5.0 and 12.0 Gyr
since the start of the star formation. The PEGASE2 models also assume a
Salpeter IMF and $z$=0.002 ($Z$\sunn/10). $M$(\HI) is taken to be 
0.75 $M_{\rm gas}$, and $L_{\rm B}$ is estimated from $M$(stars), 
with $M/L$=$\Upsilon$($B,B-V$) according to Zibetti et al. (2009).
Squares show positions of the two Lynx-Cancer triplet dwarfs. Circles denote
the three VGS galaxies and triangles the sample of  gas-rich
galaxies listed in Tab.~\ref{tab:mostgasrich}.
}
\end{figure}

Next, we use star formation models to constrain the evolutionary
history of these extremely gas rich galaxies.  For 5 of these galaxies
we only have the $(B-V)$ colour available, and we hence use this colour
to constrain the star formation history. Unfortunately, due to 
the degeneracy of the evolutionary tracks for continuous and instantaneous 
star formation laws in $B-V, V-R$ diagrams one can not distinguish between 
$\sim$3~Gyr-old constant star formation rate and $\sim$1~Gyr-old burst. 
Including $u$ or $U$ band data in the analysis allows one in many cases to do 
this. For example Pustilnik et al. (2008, 2010, 2011b) use SDSS $u,g,r$ colours
for similar gas-rich metal-poor blue galaxies to show that in most cases 
a continuous star formation law is preferable. Assuming that the
LSB galaxies in the current sample have been forming stars at a constant 
rate, we used PEGASE2 models to compute the dependence $M$(\HI)/$L_{\rm B}$
and
the $B-V$ colour on the age of the galaxy. We show in Fig.~\ref{fig:pegase}
a grid of models with $f_{\rm gas}$=0.5, 0.8, 0.9, 0.95, 0.98 and 0.99,
and star formation which started 0.5, 1, 3, 5, 12 Gyr ago. Consistent
with the mass to light ratios that we have adopted in Tab.~\ref{tab:param}
one can see that models with $f_{gas} \lesssim 0.95$ do not match the observed 
$M$(\HI)/$L_{\rm B}$ and $B-V$ colours. Further, for the ``blue''\, colours 
($B-V <$ +0.25), typical of the  gas-rich LSBDs considered here, 
the corresponding ages are $<$~3~Gyr. The conclusion is that for a continuous 
star formation law all such blue gas-rich galaxies should have  
$f_{\rm gas} \gtrsim$0.98 and should have started their star formation 
relatively recently. These implied young ages are consistent with the 
conclusions of \citet{pustilnik08,pustilnik10, pustilnik11b} that the bulk 
of stellar populations in several very metal-poor dwarfs in the Lynx-Cancer
void is relatively young.

\begin{table*}
\caption{Main parameters of galaxies with the highest $M$(\HI)/$L_{\rm B}$ ratios}
\label{tab:mostgasrich}
\begin{tabular}{lcccccc} \\ \hline
Parameter                           &  And IV                 & SBS~0335--052W   & HI 1225+01~SW   & HI 1225+01~NE   & UGCA~292        & DDO~154               \\ \hline
R.A.(J2000.0)                       & 00 42 32.30             & 03 37 38.40      & 12 26 55.00     & 12 27 46.29     & 12 38 44.63     & 12 54 05.20           \\
DEC.(J2000.0)                       & $+$40 34 18.7           & $-$05 02 36.4    & $+$01 24 35.0   & $+$01 35 57.1   & $+$32 45 01.5   & $+$27 08 59.0         \\
$B_{\rm tot}$                       & 16.56                   & 19.14            & --              & 16.0            & 16.10           & 13.94                 \\
V$_{\rm hel}$(H{\sc i})(\kms)       & 234                     & 4017             & 1226            & 1299            & 308             & 374                   \\
Distance (Mpc)                      & 6.1                     & 53.6             & 20.0            & 20.0            & 3.61            & 4.0                   \\
M$_{\rm B}^0$                       & --12.6                  & --14.70          & $>$--11.5       & --15.49         & --11.76         & --14.13               \\
  $(B-V)^0$                         & 0.20:                   & --0.12           &  --             & +0.10           & +0.08           & --0.18                \\
$\mu_{\rm B}^0$(mag~arcsec$^{-2}$)  & 23.3                    & 22.5            & $>$27           & $\sim$23.1       & 27.4            & 23.2                   \\
 12+$\log$(O/H)                     & 7.50                    & 7.00             & --              & 7.63            & 7.30            & 7.54                  \\
H{\sc i} int.flux$^{(6)}$           & 18.0                    & 0.86             & 9.1             & 23.1            & 17.6            & 82.1                  \\
$M$(\HI) (10$^{7} M_{\odot}$)       & 15.8                    & 58.3             & 86              & 220             & 5.4             & 31.0                  \\
$M$(\HI)/L$_{\rm B}$                & 13.0                    & 4.9              & $>$155          & 10.1            & 6.93            & 4.52                  \\
$f_{\rm gas}$                       & 0.99                    & 0.995            & $>$0.999        & 0.993           & 0.993           & 0.997                 \\
\hline
\multicolumn{7}{p{15.2cm}}{%
(1) -- from NED; (2) -- data for And~IV: Ferguson et al. 2000, Chengalur et
al. 2008, Pustilnik et al. 2008; (3) -- data for SBS~0335--052W: Pustilnik
et al. 2004, Ekta et al. 2009; Izotov et al. 2009; (4) --  data for
HI 1225+01~SW and HI 1225+01~NE: Salzer et al. 1991, Chengalur et al. 1995;
Turner \& MacFadyen 1997; we adopt $D$=20~Mpc, while $\sim$10 and $\sim$15
Mpc are possible alternatives; (5) -- data for UGCA~292: van Zee 2000,
Makarova et al. 1998;
(6) -- data for DDO 154: Carignan \& Freeman 1988, Carignan \& Beaulieu 1989;
Walter et al. 2008 (THINGS), O/H from Moustakas
et al. 2010.
}
\end{tabular}
\end{table*}

As discussed in the introduction, one possible solution to the discrepancy
between the number of dwarfs predicted by numerical solutions and the 
observed number of dwarfs in voids, is that the void dwarfs are fainter
than the faintest levels probed by the current surveys.  The ongoing blind 
\HI-survey ALFALFA (Haynes et al. 2011) has a significantly higher 
sensitivity and angular resolution than previous surveys. However, even in the
ALFALFA survey, objects like the faintest member of our triplet would
be difficult to detect outside the Local Volume. J0723+3624 with 
$F$(\HI)=0.48~Jy~\kms would fall below the survey detection limit were
it to be placed $\sim$1.5 times further than its current distance. ALFALFA is
hence best placed to find galaxies such as this in voids with 
$D_{\rm centre} \lesssim $20~Mpc. Detection of substructure in systems like
this triplet would however require follow up synthesis imaging observations.
We note in this context that \citet{kreckel12} find that void dwarfs show
similar small scale clustering as dwarfs in denser environments. 
For distances $D > $16~Mpc (distance moduli $\mu > $31~mag) these faint
LSBDs with $B_{\rm tot} > 19$ will not be easily identified, either via
blind \HI-surveys, nor via recently conducted wide-field spectral surveys,
like the SDSS or 2dFGRS. The existence in voids of
'unknown' population of very gas-rich LSB dwarfs with $\mb \gtrsim$ --11
which escaped detection in previous studies, hence remains a viable option.
Systematic searches for such faint dwarfs will require the next 
generation optical and radio surveys.

\section{Summary and conclusions}
\label{sec:conc}

In summary we report the discovery of an extremely gas rich triplet of
galaxies near the centre of the nearby Lynx-Cancer void. The triplet
consists of the LSB galaxies J0723+3621 (\mb = --14.2), J0723+3622
(\mb = --11.9) and J0723+3624 (\mb = --9.7) which  lie
within a projected separation of $\sim$25~kpc, and a radial velocity interval 
of $\sim$55 \kms. The $M$(\HI)/$L_{\rm B}$ ratios are $\sim$2.9, 10 and 25
respectively. All of the galaxies lie at the extreme gas rich end of the 
dwarf galaxy population. The faintest galaxy in the triplet J0723+3624 is 
one of the most gas rich galaxy known. The large 
$M$(\HI)/$L_{\rm B}$ and blue colours of the fainter two members of this 
triplet are consistent with star formation that started relatively recently 
($\lesssim$3~Gyr ago).  All three galaxies deviate significantly from
the Tully-Fisher relation, but follow the baryonic Tully-Fisher relation.
For these extremely gas rich galaxies, the baryonic mass can be accurately,
determined. We find that the baryon fraction (computed assuming they have
dark matter halos with structures as predicted by $\Lambda$CDM models)
of all of the galaxies are significantly smaller than the cosmic baryon
fraction. For the largest galaxy in the triplet, this is in contradiction
to numerical simulations which predict that it should have a baryon 
fraction comparable to the cosmic mean value. The discovery of this
very unusual dwarf triplet, along with the recent discovery by 
\citet{kreckel12} of other faint gas rich dwarfs in voids, lends
support to the suggestion that the void environment is a fertile hunting
ground for unusual, less evolved, galaxies, which in many ways resemble
high redshift young galaxies.

\section*{Acknowledgements}

SAP acknowledges the support of this work through the RFBR grants No.
10-02-92650-IND and 11-02-00261 and the Federal Target Innovative Program
under the contract No.14.740.11.0901. The authors thank Y.Lyamina for
providing the independent photometry for the faint galaxies prior
publication. This paper used observations made using the GMRT, which
is operated by NCRA-TIFR. The authors acknowledge the spectral and 
photometric data and related
information available in the SDSS database used for this study.
The Sloan Digital Sky Survey (SDSS) is a joint project of the University of
Chicago, Fermilab, the Institute for Advanced Study, the Japan Participation
Group, the Johns Hopkins University, the Max-Planck-Institute for Astronomy
(MPIA), the Max-Planck-Institute for Astrophysics (MPA), New Mexico State
University, Princeton University, the United States Naval Observatory, and
the University of Washington.
This research has made use of the NASA/IPAC Extragalactic
Database (NED).

\bsp

\label{lastpage}


\begin{thebibliography}{99}

\bibitem[\protect\citeauthoryear{Abazajian et al.}{2009}]{DR7}
       Abazajian K.N., Adelman-McCarthy J.K., Ag\"ueros M.A. et al.,
	2009, ApJS, 182, 543
\bibitem[\protect\citeauthoryear{Begum et al.}{2008a}]{begum08a}
   Begum A., Chengalur J.N., Karachentsev I.D., Sharina M.E.,
    2008a, MNRAS, 386, 138
\bibitem[\protect\citeauthoryear{Begum et al.}{2008b}]{begum08}
   Begum A., Chengalur J.N., Karachentsev I.D., Sharina M.E., Kaisin S.S.,
    2008b, MNRAS, 386, 1667
\bibitem[\protect\citeauthoryear{Blanton et al.}{2005}]{blanton05}
   Blanton M.B., Lupton R., Schlegel D.J., Strauss M.A., Brinkmann J.,
   Fukugita M., Loveday J., 2005, ApJ, 631, 208
\bibitem[\protect\citeauthoryear{Carignan \& Freeman}{1988}]{Carignan88}
    Carignan C. \& Freeman K.C., 1988, ApJ, 332, L33
\bibitem[\protect\citeauthoryear{Carignan \& Beaulieu}{1989}]{Carignan89}
    Carignan C. \& Beaulieu S., 1989, ApJ, 347, 760
\bibitem[\protect\citeauthoryear{Chengalur et al.}{1995}]{Chengalur95}
    Chengalur J.N., Giovanelli R., Haynes M.P., 1995, AJ, 109, 2415
\bibitem[\protect\citeauthoryear{Chengalur et al.}{2008}]{Chengalur08}
    Chengalur J.N., Begum A., Karachentsev I.D., Sharina M., Kaisin S., 2008,
    Galaxies in  the Local Volume, Astrophysics and Space Science Proceedings,
    Volume. ISBN 978-1-4020-6932-1. Springer Netherlands, 2008, p.65
    (arXiv:0711.2153)
\bibitem[\protect\citeauthoryear{Davis et al.}{1985}]{davis85} 
  Davis M., Efstathiou G., Frenk C.~S., White S.~D.~M., 1985, 
  ApJ, 292, 371 
\bibitem[\protect\citeauthoryear{De Rijcke et al.}{2007}]{derijcke07} 
  De Rijcke S., Zeilinger W.~W., Hau G.~K.~T., Prugniel P., 
  Dejonghe H., 2007, ApJ, 659, 1172 
\bibitem[\protect\citeauthoryear{Ekta, Pustilnik, \& Chengalur}
  {Ekta et al}{2009}]{Ekta09}
  Ekta B., Pustilnik S.A., Chengalur J.N., 2009, \mnras, 397, 963
\bibitem[\protect\citeauthoryear{Ferguson, Gallagher, \& Wise}{2000}]
   {Ferguson00}
   Ferguson A.M.N., Gallagher J.S., \& Wise R.F.G., 2000, AJ, 120, 821
\bibitem[\protect\citeauthoryear{Geha et al.}{2006}]{geha06}
	Geha M., Blanton M.R., Masjedi M., West A.A., 2006, ApJ, 653, 240
\bibitem[\protect\citeauthoryear{Geller \& Huchra}{1989}]{geller89}
  Geller M.~J., Huchra J.~P., 1989, Sci, 246, 897 
\bibitem[\protect\citeauthoryear{Gottl{\"o}ber et al.}{2003}]{gottlober03} 
  Gottl{\"o}ber S., {\L}okas E.~L., Klypin A., Hoffman Y., 2003, 
  MNRAS, 344, 715 
\bibitem[\protect\citeauthoryear{Haynes et al.}{2011}]{ALFALFA.4}
	Haynes M.P., Giovanelli R., Martin A.M., et al. 2011,
	AJ, 142, 170
\bibitem[\protect\citeauthoryear{Hoeft and Gottl\"ober}{2010}]{hoeft10}
   Hoeft M., Gottl\"ober S., 2010, Advances in Astronomy, v.2010,       â
  Article ID 693968, 16 pp.
\bibitem[\protect\citeauthoryear{Hoeft et al.}{2006}]{hoeft06}
     Hoeft M., Yepes G., Gottl\"ober S., Springel V., 2006, MNRAS, 371, 401
\bibitem[\protect\citeauthoryear{Huchtmeier, Hopp, \& Kuhn}{1997}] 
        {huchtmeier97} Huchtmeier W.~K., Hopp U., Kuhn B., 1997, 
        A\&A, 319, 67 
\bibitem[\protect\citeauthoryear{Izotov et al.}{2009}]{Izotov09}
	Izotov Y.I., Guseva N.G., Fricke K.J., Papaderos P., 2009,
	A\&A, 503, 61
\bibitem[\protect\citeauthoryear{Jarosik et al.}{2011}]{jarosik11} 
  Jarosik N., et al., 2011, ApJS, 192, 14 
\bibitem[\protect\citeauthoryear{J\"oeveer, Einasto \& Tago}{1978}]{joeveer78}
	J\"oeveer, M., Einasto J., \& Tago E.,  2078, MNRAS, 185, 357
\bibitem[\protect\citeauthoryear{Karachentsev et al.}{2003}]{Kara2003}
	Karachentsev I.D., Sharina M.E., Dolphin A.E., et al., 2003,
	A\&A, 398, 467
\bibitem[\protect\citeauthoryear{Kirshner et al.}{1981}]{kirshner81} 
  Kirshner R.~P., Oemler A., Jr., Schechter P.~L., Shectman S.~A., 
  1981, ApJ, 248, L57 


\bibitem[\protect\citeauthoryear{Kreckel et al.}{2011}]{kreckel11}
    Kreckel K., Joung M.R., Cen R.,      2011, ApJ, 735, 132

\bibitem[\protect\citeauthoryear{Kreckel et al.}{2012}]{kreckel12}
    Kreckel K.,  Platen E., Aragon-Calvo M.A., van Gorkom J.H., van de
    Weygaert R., van der Hulst J.M., Beygu B., 2012, AJ, 144, 16

\bibitem[\protect\citeauthoryear{Lindner et al.}{1996}]{lindner96}
  Lindner U., Einasto M., Einasto J., et al.,
     1996, A\&A, 314, 1

\bibitem[\protect\citeauthoryear{Lupton et al.}{2001}]{Lupton01}
     Lupton R., Gunn J.E., Ivezi\'c Z. et al., 2001,
     in: Harnden F.R., Jr., Primini F.A. , \& Payne H.E., eds,
     Astronomical Data Analysis Software and Systems X, ASP Conf.
     Ser. 238, Astron. Soc. Pac., San Francisco, p. 269

\bibitem[\protect\citeauthoryear{Lupton et al.}{2005}]{Lupton05}
   \mbox{Lupton~R.,~et~al.~2005},
  http://www.sdss.org/dr5/algorithms\\/sdssUBVRITransform.html\#Lupton2005
\bibitem[\protect\citeauthoryear{Makarova et al.}{1998}]{makarova98} 
  Makarova L., Karachentsev I., Takalo L.~O., Hein\"am\"aki P.,
  Valtonen M., 1998, A\&AS, 128, 459 
\bibitem[\protect\citeauthoryear{McGaugh \& de Blok}{1997}]{mcgaugh97} 
  McGaugh S.~S., de Blok W.~J.~G., 1997, ApJ, 481, 689 
\bibitem[\protect\citeauthoryear{McGaugh et al.}{2010}]{mcgaugh10} 
  McGaugh S.~S., Schombert J.~M., de Blok W.~J.~G., Zagursky M.~J., 
  2010, ApJ, 708, L14 
\bibitem[\protect\citeauthoryear{Moustakas et al.}{2010}]{Moustakas10}
      Moustakas J., Kennicutt R.C., Tremonti C.A., Dale D.A.. Smith J.-D.T.,
      Calzetti D., 2010, ApJS, 190, 233.
\bibitem[\protect\citeauthoryear{Patiri et al.}{2006}]{patiri06}
      Patiri S.G., Prada F., Holtzman J., Klypin A., Betancort-Rijo J.,
     2006, MNRAS, 372, 1710
\bibitem[\protect\citeauthoryear{Park et al.}{2007}]{park07} 
  Park C., Choi Y.-Y., Vogeley M.~S., Gott J.~R., III, Blanton M.~R., SDSS 
  Collaboration, 2007, ApJ, 658, 898 
\bibitem[\protect\citeauthoryear{Peebles}{2001}]{peebles01}
      Peebles P.J.E.,  2001, ApJ, 557, 459
\bibitem[\protect\citeauthoryear{Postman \& Geller}{1984}]{postman84} 
  Postman M., Geller M.~J., 1984, ApJ, 281, 95 
\bibitem[\protect\citeauthoryear{Prasad \& Chengalur}{2012}]{prasad12} 
  Prasad J., Chengalur J., 2012, ExA, 33, 157 
\bibitem[\protect\citeauthoryear{Pustilnik et al.}{1995}]{pustilnik95}
  Pustilnik S.A., Ugryumov A.V., Lipovetsky V.A., Thuan T.X.,  Guseva N.G.,
  1995, ApJ, 443, 499
\bibitem[\protect\citeauthoryear{Pustilnik et al.}{2001}]{pustilnik01} 
  Pustilnik S.A., Brinks E., Thuan T.X., Lipovetsky V.A.,
  Izotov Y.~I., 2001, AJ, 121, 1413 
\bibitem[\protect\citeauthoryear{Pustilnik et al.}{2002}] {pustilnik02} 
  Pustilnik S.~A., Martin J.-M., Huchtmeier W.~K., Brosch N., 
  Lipovetsky V.~A., Richter G.~M., 2002, A\&A, 389, 405 
\bibitem[\protect\citeauthoryear{Pustilnik et al.}{2003}]{SAO0822}
  Pustilnik S.A, Kniazev A.Y., Pramsky A.G., Ugryumov A.V.,
  Masegosa J.,  2003, A\&A, 409, 917
\bibitem[\protect\citeauthoryear{Pustilnik, Pramskij, Kniazev}{2004}]{SBS0335}
  Pustilnik S.A, Pramskij A.G.,  Kniazev A.Y.,  2004, A\&A, 425, 51
\bibitem[\protect\citeauthoryear{Pustilnik, Kniazev \& Pramskij}{Pustilnik
    et al.}{2005}]{pustilnik05} Pustilnik S.A., Kniazev A.Y., Pramskij A.G.,
  2005, \aap, 443, 91
\bibitem[\protect\citeauthoryear{Pustilnik et al.}{2008}]{pustilnik08}
  Pustilnik S.A., Tepliakova A.L., Kniazev A.Y., Burenkov A.N.,
  2008, Astrophys. Bulletin, 63, 102 (arXiv:0712.4205)
\bibitem[\protect\citeauthoryear{Pustilnik et al}{2010}]{pustilnik10}
  Pustilnik S.A., Tepliakova A.L., Kniazev A.Y., Martin J.-M., Burenkov A.N.,
  2010, MNRAS, 401, 333
\bibitem[\protect\citeauthoryear{Pustilnik \& Tepliakova}{2011}]{pustilnik11a}
 Pustilnik S.A., Tepliakova A.L., 2011, MNRAS, 415, 1188  (Paper~I)
\bibitem[\protect\citeauthoryear{Pustilnik, Tepliakova \& Kniazev}{2011}]
  {pustilnik11b} Pustilnik S.A., Tepliakova A.L., Kniazev A.Y., 2011,
  Astrophys.~Bulletin, 66, 255 (Paper~II)  (arXiv:1108.4850)
\bibitem[\protect\citeauthoryear{Pustilnik et al.}{2011c}]{pustilnik11c}
   Pustilnik S.A., Martin J.-M., Tepliakova A.L., Kniazev A.Y.,
   2011, MNRAS, 417, 1335  (Paper~III)
\bibitem[\protect\citeauthoryear{Roberts}{1969}]{Roberts69}
    Roberts M.S., 1969, AJ, 74, 859
\bibitem[\protect\citeauthoryear{Rojas et al.}{2005}]{rojas05} 
  Rojas R.~R., Vogeley M.~S., Hoyle F., Brinkmann J., 2005, ApJ, 624, 571 
\bibitem[\protect\citeauthoryear{Rojas et al.}{2004}]{rojas04} 
  Rojas R.~R., Vogeley M.~S., Hoyle F., Brinkmann J., 2004, ApJ, 617, 50 
\bibitem[\protect\citeauthoryear{Salzer et al.}{1991}]{Salzer91}
    Salzer J.J., di Serego Alighieri S., Matteicci F., Giovanelli R.,
    Haynes M.P., 1991, AJ, 101, 1258
\bibitem[\protect\citeauthoryear{Salzer}{1992}]{salzer92} Salzer 
  J.~J., 1992, AJ, 103, 385 
\bibitem[\protect\citeauthoryear{Santiago et al.}{1995}]{santiago95} 
  Santiago B.~X., Strauss M.~A., Lahav O., Davis M., Dressler A., 
  Huchra J.~P., 1995, ApJ, 446, 457 
\bibitem[\protect\citeauthoryear{Schlegel, Finkbeiner, Douglas}{1998}]
   {Schlegel98}
    Schlegel D.J., Finkbeiner D.P., Douglas M., 1998, ApJ, 500, 525
\bibitem[\protect\citeauthoryear{Tikhonov \& Klypin}{2009}]{tikhonov09} 
  Tikhonov A.~V., Klypin A., 2009, MNRAS, 395, 1915 
\bibitem[\protect\citeauthoryear{Tinker \& Conroy}{2009}]{tinker09}
  Tinker J.L., \& Conroy C., 2008, ApJ, 691, 633
\bibitem[\protect\citeauthoryear{Tully \& Pierce}{2000}]{tully00}
    Tully R.B., Pierce M.J., 2000, ApJ, 533. 744
\bibitem[\protect\citeauthoryear{Tully et al.}{2008}]{Tully08}
    Tully R.B., Shaya E.J., Karachentsev I.D., Courtois H.M., Kocevski D.D.,
    Rizzi L., Peel A.,   2008, \apj, 676, 184

\bibitem[\protect\citeauthoryear{Turner \& MacFadyen}{1997}]{Turner97}
    Turner N.J.J., MacFadyen A., 1997, MNRAS, 285, 125

\bibitem[\protect\citeauthoryear{van Zee}{2000}]{vanZee00}
      van Zee L., 2000, ApJ, 543, L31
\bibitem[\protect\citeauthoryear{Verheijen \& Sancisi}{2001}] {verheijen01} 
  Verheijen M.~A.~W., Sancisi R., 2001, A\&A, 370, 765 

\bibitem[\protect\citeauthoryear{Walter et al.}{2009}]{Walter08}
      Walter F., Brinks E., de Block W.J.G., Bigiel F., Kennicutt R.C.,
      Thornley M.D., Leroy A., 2008, AJ, 136, 2563

\bibitem[\protect\citeauthoryear{White et al.}{1987}]{white87} 
  White S.~D.~M., Davis M., Efstathiou G., Frenk C.~S., 1987, 
  Nature, 330, 451 

\bibitem[\protect\citeauthoryear{Zackrisson, Bergvall \& \"Ostlin}{2005}]
    {Zackrisson05}
       Zackrisson E., Bergvall N., \"Ostlin G., 2005, A\&A, 435, 29

\bibitem[\protect\citeauthoryear{Zibetti, Charlot \& Rix}{Zibetti et al.}
       {2009}]{Zibetti09}
       Zibetti S., Charlot S. \& Rix H.-W., 2009, MNRAS, 400, 1181


\end{thebibliography}
\end{document}